\documentclass{elsart}

\usepackage{amssymb}

\begin{document}

\begin{frontmatter}



\title{Binary Bargmann Symmetry Constraints of Soliton Equations}


\author{Wen-Xiu Ma}

\address{Department of Mathematics, City University of Hong Kong, 
Hong Kong, Kowloon, China}

\begin{abstract}

Binary Bargmann symmetry constraints are applied to 
decompose soliton equations into 
finite-dimensional Liouville integrable Hamiltonian systems, generated from 
so-called constrained flows.
The resulting constraints on the potentials of soliton equations
give rise to involutive solutions to soliton equations,
and thus the integrability by quadratures are shown for soliton equations
by the constrained flows.
The multi-wave interaction equations associated with the $3\times 3$ matrix 
AKNS spectral problem are chosen as an illustrative example
to carry out binary Bargmann symmetry constraints. 
The Lax representations and the corresponding ${\bf r}$-matrix formulation are established
for the constrained flows of the multi-wave interaction equations, and 
the integrals of motion generated 
from the Lax representations are utilized to 
show the Liouville integrability for the resulting constrained flows.
Finally, involutive solutions 
to the multi-wave interaction equations are presented.

\end{abstract}

\end{frontmatter}

\setlength{\parindent}{15pt}

\def \part {\partial}
\def \be {\begin{equation}}
\def \ee {\end{equation}}
\def \bea {\begin{eqnarray}}
\def \eea {\end{eqnarray}}
\def \ba {\begin{array}}
\def \ea {\end{array}}
\def \si {\sigma}
\def \al {\alpha}
\def \la {\lambda }
\def\D{\displaystyle}
\def\diag {\textrm{diag}}

\newcommand{\Z}{\mathbb{Z}}
\newcommand{\C}{\mathbb{C}}
\newcommand{\R}{\mathbb{R}}

\newtheorem{lemma}{Lemma}[section]
\newtheorem{theorem}{Theorem}[section]
\newtheorem{definition}{Definition}[section]
\newtheorem{proposition}{Proposition}[section]

\renewcommand{\theequation}{\thesection.\arabic{equation}}

\section{Introduction}
\setcounter{equation}{0}

The nonlinearization technique 
has been attracting broad attention in the soliton community
\cite{Cao-SC1990,CaoG-book1990,ZengL-AMS1990}. 
It provides a powerful approach for analyzing 
soliton equations, in both continuous and discrete cases, especially for showing the integrability by quadratures for soliton equations.
A kind of specific Bargmann symmetry constraints \cite{MaS-PLA1994}
yields the required constraints on the potentials of 
soliton equations in nonlinearization. One of two conserved covariants involved in 
each Bargmann symmetry constraint is Lie point, but the other 
is not Lie point, contact or Lie-B\"acklund, which 
is expressed in terms of eigenfunctions and adjoint 
eigenfunctions of spectral problems associated with soliton equations.
Such symmetry constraints nonlinearize spectral problems of soliton equations
into finite-dimensional Liouville integrable systems,
and the constraints on the potentials present
involutive solutions of soliton equations.

Binary Bargmann symmetry constraints also establish a bridge between
infinite-dimensional soliton equations and
finite-dimensional Liouville integrable systems 
\cite{MaS-PLA1994}. The  
constrained flows resulted from 
spectral problems of soliton equations
are utilized to show the integrability by quadratures 
for soliton equations \cite{Ma-PA1995,MaFO-PA1996}
and pave a method of separation of variables  
for soliton equations \cite{EilbeckEKT-JPA1994,MaZ-krustal2000}.
The study of binary symmetry constraints itself is an important part of
the kernel of the mathematical theory of nonlinearization.

This paper is structured as follows.
In section 2, a general theory is recalled for reference. 
Then starting from section 3, an illustrative example 
is carried out to manipulate binary Bargmann symmetry constraints.
Finally in section \ref{sec:finalsectionofitaly00},
a summary is given along with concluding remarks.

\section{General theory}
\setcounter{equation}{0}

Let us address a skeleton theory of 
binary Bargmann symmetry constraints
(for example, see \cite{MaF-book1996} for details).
We start from a matrix spectral problem  
\begin{equation}
 \phi_x=U\phi = U(u,\lambda)\phi,\ U=(U_{ij})_{r\times r},\ 
  \phi=(\phi_1,\cdots,\phi_r)^T
\label{gsp}
\end{equation}
with a spectral parameter $\la $ and a potential $u=(u_1,\cdots,u_q)^T$.
Suppose that the compatability conditions
$ U_{t_m}-V_x^{(m)}+[U,V^{(m)}]=0,\ m\ge 0,$
 of the spectral problem (\ref{gsp})
and the associated spectral problems
\be 
\phi_{t_m}=V^{(m)}\phi = V^{(m)}(u,u_x,\cdots;\lambda)\phi,\ V^{(m)}=(V^{(m)}_{ij})_{r\times r },\ m\ge 0, \label{gassosp}
\ee  
determine an isospectral ($\la _{t_m}=0$) soliton hierarchy
\be 
u_{t_m}=X_m(u)=JG_m=J\frac{\delta {\tilde H_m}}{\delta u},\ m\ge 0,
\label{sh}
\ee 
where the Hamiltonian operator $J$ and the Hamiltonian functionals ${\tilde H}_m$ 
can be determined by a trace identity.
Evidently, the compatability conditions of 
the adjoint spectral problem
and the adjoint associated spectral problems 
\be 
\psi _x=-U^T(u,\la )\psi ,\ 
\psi _{t_m} =-V^{(m)T}(u,\la )\psi,\ 
m\ge 0,
\ee 
where $\psi =(\psi _1,\cdots ,\psi_r)^T$, 
still give rise to the same soliton hierarchy (\ref{sh}).

It has been pointed out \cite{MaS-PLA1994} that 
$
J\D\frac {\delta \lambda }{\delta u}=E^{-1} J\psi ^T
\frac{\part U(u,\la )}{\part u}\phi
$ 
is a common 
symmetry of all equations in the hierarchy (\ref{sh}),
where $E$ is the normalized constant.
Upon introducing 
$N$ distinct eigenvalues $\la _1,\cdots,\la _N$,
we obtain
\be 
\phi^{(s)}_x=U(u,\la _s)\phi^{(s)}, \ \psi^{(s)}_x=-U^T(u,\la _s)\psi^{(s)},
\ 1\le s\le N;\qquad 
\label{gxpartofcf}
\ee
\be   
\phi^{(s)}_{t_m}=V^{(m)}(u,\la _s)\phi^{(s)}, \ \psi^{(s)}_{t_m}
=-V^{(m)T}(u,\la _s)\psi^{(s)},\ 1\le s\le N,
\label{gtpartofcf}
\ee 
where the corresponding eigenfunctions and adjoint eigenfunctions 
are denoted by $\phi^{(s)}$ and $\psi^{(s)}$, $1\le s\le N$.
It is assumed that the conserved covariant 
$G_{m_0}$ is Lie point, and then
the so-called binary Bargmann symmetry constraint reads 
\be
X_{m_0}=\sum_{s=1}^N E_s
J\frac {\delta \la _s}{\delta u},\ \textrm{i.e.,} \ 
JG_{m_0}=
J\sum_{s=1}^N\psi^{(s)T}\frac {\part U(u,\la _s)}{\part u}\phi^{(s)},
\label{gsy} \ee 
where 
$E_s,\ 1\le s\le N,$ are $N$ normalized constants.
The right-hand side of the symmetry constraint (\ref{gsy}) is a linear combination of $N$ symmetries 
$
E_sJ\D{\delta \lambda_s }/{\delta u},\ 1\le s\le N$.
Such symmetries 
(or the corresponding conserved covariants $\D{\delta \lambda_s }/{\delta u},\ 1\le s\le N$)
are not Lie point, contact or Lie-B\"acklund
symmetries, since $\phi^{(s)}$ and $\psi^{(s)}$ can not be expressed 
in terms of $x$, $u$ and derivatives of $u$ with respect to $x$ to some finite order. 

Suppose that the symmetry constraint (\ref{gsy}) has an inverse function
\be u=\widetilde u(\phi^{(1)},\cdots,\phi^{(N)};\psi^{(1)},\cdots,\psi^{(N)}).
\ee 
Replacing $u$ with $\widetilde u$ in 
the system ({\ref{gxpartofcf}) or 
the system (\ref{gtpartofcf}), we obtain the so-called spatial constrained flow and 
the so-called temporal
constrained flow:
\be 
\phi^{(s)}_x=U(\widetilde u,\la _s)\phi^{(s)}, \ \psi^{(s)}_x=-U^T(\widetilde u,\la _s)\psi^{(s)},
\ 1\le s\le N;\qquad 
\label{xpartofcf}
\ee 
\be 
\phi^{(s)}_{t_m}=V^{(m)}(\widetilde u,\la _s)\phi^{(s)}, \ \psi^{(s)}_{t_m}
=-V^{(m)T}(\widetilde u,\la _s)\psi^{(s)}, \ 1\le s\le N.
\label{tpartofcf}
\ee 
The former is a system of ordinary differential equations,
and the latter is usually a system of 
partial differential equations but it can be transformed
into a system of ordinary differential equations under the control of (\ref{xpartofcf}).

The main question of symmetry constraints is to show that 
the spatial constrained flow (\ref{xpartofcf})
and the temporal constrained flows (\ref{tpartofcf}) 
under the control of (\ref{xpartofcf})
are Liouville integrable. 
Then if $\phi^{(s)}$ and $\psi ^{(s)}$, $1\le s\le N$,
solve two constrained flows (\ref{xpartofcf}) and (\ref{tpartofcf}) simultaneously,  
 $u=\widetilde u$
will present a solution to 
the $m$th soliton equation $u_{t_m}=X_m(u)$. It also follows that the soliton equation 
$u_{t_m}=X_m(u)$ is decomposed into two finite-dimensional Liouville integrable 
systems, and   
the constraint $u=\widetilde u$ gives a B\"acklund transformation between soliton equations and 
the resulting finite-dimensional Liouville integrable 
systems \cite{MaG-book1999}.
The whole process to carry out binary Bargmann symmetry constraints 
is called binary nonlinearization \cite{Ma-CAM1997,Ma-PA1995}.

\section{Multiwave interaction equations}
\setcounter{equation}{0}

Let us begin with the 3$\times $3 matrix AKNS spectral problem: 
\be \phi _x=U\left( u,\lambda \right) \phi ,\ U=\left( 
\begin{array}{ccc}
\al _1\lambda  & u_{12} & u_{13}  \\ 
u_{21} & \al _2 \la & u_{23}  \\ 
u_{31} & u_{32} & \al _3\la   
\end{array}
\right)=\lambda U_0  +U_1 ,\ \phi =\left( 
\begin{array}{c}
\phi _1 \\ 
\phi _2 \\ 
\phi _3 
\end{array}
\right), 
\ee 
where $U_0=\textrm{diag}(\al _1,\al _2,\al _3)$, and $\al _1,\al _2,\al _3$ 
are distinct constants, and the potential $u$ 
is defined by 
\be u=\rho (U_1)=(u_{21},u_{12},u_{31},u_{13},u_{32},u_{23})^T .
\label{eq:defofroh}\ee 
We introduce an
associated spectral problem: 
\[\phi _t= V\left( u,\lambda \right) \phi ,\ V=\left( 
\begin{array}{ccc}
\beta _1\lambda  & v_{12} & v_{13}  \\ 
v_{21} & \beta _2 \la & v_{23}  \\ 
v_{31} & v_{32} & \beta _3\la   
\end{array}
\right)=\lambda V_0 +V_1 ,
\]
where $V_0=\textrm{diag}(\beta _1,\beta _2,\beta _3)$, and $\beta _1,\beta _2,\beta _3$ 
are distinct constants.
Under the isospectral condition $\la _t=0$, the compatibility condition
$ U_{t}-V_x+[U,V]=0$ is equivalent to 
\be U_{1t}-V_{1x}+[U_1,V_1]=0,\ [U_0,V_1]=[V_0,U_1].
\label{eq:matrixformof3MWIEs}
\ee 
They give rise to the multi-wave interaction equations 
\be 
u_{ij,t}=\frac{\beta_i-\beta_j}{\alpha_i-\alpha_j}u_{ij,x}
+\sum^3_{\stackrel{k=1}{k\not=i,j}}(\frac{\beta_k-\beta_i}{\alpha_k-\alpha_i}
-\frac{\beta_k-\beta_j}{\alpha_k-\alpha_j})u_{ik}u_{kj},\  1\leq i\ne j\leq 3, \ \,
\label{eq:3MWIEs}\ee 
which contain three-wave interaction equations
arising in fluid dynamics and plasma physics 
\cite{ZakhrovM-SPJEPTL1973,Kaup-SAM1976}, 
with $U_1$ being chosen to be an anti-Hermitian matrix. 
Note that the compatability condition of
the adjoint spectral problem
and adjoint associated spectral problem: 
\be \psi_x=-U^T(u,\la )\psi ,\ \psi _t=-V^T(u,\la )\psi,\ \psi=(\psi _1,\psi_2,\psi_3)^T \ee
still gives rise to the above multi-wave interaction equations.
It is easy to find that the multi-wave interaction equations 
(\ref{eq:3MWIEs}) have the Hamiltonian structure
\be 
 u_{ij,t}=J\frac {\delta {\tilde H}}{\delta u_{ij}} ,\ 1\le i\ne j\le 3,\ee 
where the Hamiltonian operator $J$ and the Hamiltonian ${\tilde H}$
are defined by 
\bea && 
J=\textrm{diag}\Bigl( (\alpha_1-\alpha_2)\sigma_0, (\alpha_1-\alpha_3)\sigma_0,(\alpha_2-\alpha_3)\sigma_0\Bigl),\ 
\sigma_0=\left( \begin{array} {cc} 0&\, 1 \vspace{2mm} \\ -1&\, 0 \end{array}
 \right ),\ \ 
\\
&& {\tilde H}=\int H\,dx,\ H=
\frac 1 {3(\al _1 -\al _2)}\Bigl( \frac{\beta _3-\beta _2}{\al _3-\al _2}-
\frac{\beta _3-\beta _1}{\al _3-\al _1}\Bigr)
\bigl(u_{12}u_{23}u_{31}\bigr. \nonumber \\  &&
\ \ \ 
+\bigl. u_{21}u_{13}u_{32} \bigr)
+\frac{\beta _1-\beta _2}{2(\al _1-\al _2)^2}
(u_{12}u_{21,x}-u_{21}u_{12,x})
+\textrm{cycle}(1,2,3).\ 
\eea 

\section{Binary Bargmann symmetry constraints}
\setcounter{equation}{0}

Let us consider binary Lax systems of the multi-wave interaction equations
(\ref{eq:3MWIEs}):
\be  \phi _x =U(u,\lambda )\phi ,\   \psi _x=-U^T(u,\lambda )\psi ;
\ee 
\be   \phi _t =V(u,\lambda )\phi , \  
\psi _t=-V^T(u,\lambda )\psi . 
\ee 
Based on the Hamiltonian theory, 
the conserved functional $\lambda(u) $ gives rise to a symmetry of 
(\ref{eq:3MWIEs}):
\be EJ\frac {\delta \la }{\delta u}=E 
\rho([U_0,\rho^{-1}(\frac {\delta \lambda }{\delta u})])
=\rho([U_0,
 \rho^{-1}(\psi ^T\frac {\partial U(u,\lambda )}{\partial u}\phi )]),
\ee 
where $E$ is the normalized constant and $\rho $ is the mapping defined by (\ref{eq:defofroh}).
Having introduced $N$ distinct eigenvalues $\la _1,\cdots, \la _N$,
we obtain
\be  \phi^{(s)} _{x}=U(u,\la _s)
\phi ^{(s)},\  
\psi ^{(s)}_{x}=-U^T(u,\la _s)
\psi^{(s)},\ 1\le s\le N;
\label{eq:originalxpartofbinaryLaxsystems}
 \ee \be 
\phi^{(s)} _{t}=V(u,\la _s)
\phi ^{(s)},\  
\psi ^{(s)}_{t}=-V^T(u,\la _s)
\psi^{(s)},\ 1\le s\le N, \label{eq:originaltpartofbinaryLaxsystems}
\ee  
where 
$\phi^{(s)}$ and $\psi^{(s)}$ are assumed to be denoted by 
\be 
\phi^{(s)}=(\phi_{1s},\phi_{2s},\phi_{3s})^T,
\ \psi^{(s)}=(\psi_{1s},\psi_{2s},\psi_{3s})^T,\ 1\le s\le N.\ee 
This multi-eigenvalue case
yields a general symmetry of (\ref{eq:3MWIEs})
(not Lie point, contact or Lie B\"acklund type):
\be Z_0:=
\rho ([U_0,\rho^{-1}(\D \sum_{s=1}^N \psi ^{(s)T} \frac{\partial U(u,\lambda _s)}{\partial u}\phi^{(s)})])
= \rho([U_0,\D \sum_{s=1}^N \phi ^{(s)}\psi^{(s)T}]). \ee  

On the other hand, it is easy to find that 
the multi-wave interaction equations 
(\ref{eq:3MWIEs}) have a Lie point symmetry  
\be Y_0:= \rho ([\Gamma, U_1]),\ 
\Gamma=\diag (\gamma_1,\gamma_2,\gamma _3),\ \gamma_i\ne \gamma _j,\ 1\le i\ne j\le 3, \ee
where $\gamma_1,\gamma_2,\gamma_3$ are constants.
Actually, we can directly prove that $Y_0$ and $Z_0$ are two symmetries of 
the multi-wave interaction equations 
(\ref{eq:3MWIEs}). 
Set 
\be \delta U_1=[\Gamma, U_1]\ \textrm{or}\  [U_0,\sum_{s=1}^N 
\phi ^{(s)}\psi^{(s)T}],\ee
then $[U_0,\delta V_1]=[V_0,\delta U_1]$ determines
\be \delta V_1=[\Gamma, V_1]\ \textrm{or}\ [V_0,\sum_{s=1}^N
\phi ^{(s)}\psi^{(s)T}].\ee  
Now in view of (\ref{eq:matrixformof3MWIEs}),
the symmetry problem requires to show that 
\[ (\delta U_1,\delta V_1)=([ \Gamma, U_1],[\Gamma, V_1])\ \textrm{or} \ 
([U_0,\D \sum_{s=1}^N
\phi ^{(s)}\psi^{(s)T}],[V_0,\sum_{s=1}^N
\phi ^{(s)}\psi^{(s)T}])\ \] 
satisfies the linearized system of the multi-wave interaction equations 
(\ref{eq:3MWIEs}): 
\be 
 (\delta U_1)_t-(\delta V_1)_x+[\delta U_1,V_1]+[U_1,\delta V_1]=0,\ee
which just needs a direct computation. 

Therefore, binary Bargmann symmetry constraint of (\ref{eq:3MWIEs}) 
reads as
\be Y_0=Z_0,\ \textrm{i.e.,}\ 
[\Gamma, U_1]=
[U_0,\sum_{s=1}^N
\phi ^{(s)}\psi^{(s)T} ]. \label{eq:bscof3MWIEs}\ee 
Solving this for $u$, we have the constraints on the potentials
\be  u_{ij}=\tilde u_{ij}:=
\frac{\alpha_i-\alpha_j}{\gamma_i-\gamma_j}\langle\Phi_i,\Psi_j\rangle,\ 
  1\leq i\ne j\leq 3, \label{eq:defoftildeu}\ee 
where  $\langle\cdot,\cdot\rangle$ denotes the standard inner-product 
of ${\R}^N$, and 
\be \Phi_{i}=(\phi_{i1},\cdots,\phi_{iN})^T,\ 
\Psi_{i}=(\psi_{i1},\cdots,\psi_{iN})^T,\ 1\le i\le 3.
\ee

The substitution of $u$ with 
$ {\tilde u}=({\tilde u}_{21},{\tilde u}_{12},
{\tilde u}_{31},{\tilde u}_{13},
{\tilde u}_{23},{\tilde u}_{32} )$
in binary Lax systems (\ref{eq:originalxpartofbinaryLaxsystems})
and (\ref{eq:originaltpartofbinaryLaxsystems})
leads to 
the so-called constrained flows:
\be  
\phi^{(s)} _{x}=U(\tilde u ,\la _s) \phi ^{(s)},\ 
\psi^{(s)}_{x}=-U^T(\tilde u ,\la _s) \psi^{(s)},\ 1\le s\le N;\label{eq:xcfof3NWIEs}
\ee \be  \phi ^{(s)}_{t}=V(\tilde u ,\la _s) \phi ^{(s)} 
,\ \psi^{(s)}_{t}=-V^T(\tilde u ,\la _s) \psi ^{(s)},\ 1\le s\le N.  \label{eq:tcfof3NWIEs}\ee   
These are two systems of ordinary differential equations.
In the following two sections,
we would like to show their Liouville integrability.

\section{Hamiltonian structures and Lax representations}
\setcounter{equation}{0}

In order to prove that two constrained flows 
are Liouville integrable, we need generate 
their Hamiltonian structures and integrals of motion. 
First, the spatial constrained flow (\ref{eq:xcfof3NWIEs})
can be easily expressed as the Hamiltonian form
\be 
\Phi_{ix}=-\frac{\partial H^x}{\partial \Psi_i},\ \Psi_{ix}=\frac{\partial H^x}{\partial \Phi_i}
,\ 1\leq i\leq 3,
\label{eq:Hsofxpartofcfs}
\ee
with the Hamiltonian  
\be 
H^x=- \sum^3_{k=1}\alpha_k\langle A    \Phi_k,\Psi_k\rangle
-\sum_{1\leq k<l\leq 3}\frac{\alpha_k-\alpha_l}{\beta_k-\beta_l}\langle\Phi_k,\Psi_l\rangle\langle\Phi_l,\Psi_k\rangle,
\ee 
where  $
A=\textrm{diag}(\lambda _1,\lambda _2,\cdots ,\lambda _N).$
Second, the temporal constrained flow (\ref{eq:tcfof3NWIEs})
has a similar Hamiltonian structure
\be 
\Phi_{it}=-\frac{\partial H^t}{\partial \Psi_i},
\ \Psi_{it}=\frac{\partial H^t}{\partial \Phi_i},\ 1\leq i\leq 3,
\label{eq:Hsoftpartofcfs}
\ee 
with the Hamiltonian
\be 
H^t=- \sum_{k=1}^3\beta _k\langle A   \Phi _k,\Psi_k\rangle
-\sum_{1\le k<l\le 3}\frac{\beta _k-\beta_l}{\gamma _k-\gamma _l}
\langle\Phi _k,\Psi _l\rangle\langle\Phi _l,\Psi_k\rangle. \ee 
It is known that 
Lax representations are basic objects 
to generate integrals of motion.
In order to present Lax representations
of the constrained flows 
(\ref{eq:xcfof3NWIEs})
and (\ref{eq:tcfof3NWIEs}),
let us define a Lax operator $L(\la )$ by
\be 
L(\la ) =\left(\ba {ccc}
\D \gamma_1+ 
\D \sum_{l=1}^{N}
\frac{\phi_{1l}\psi_{1l}}{\lambda -\lambda _l}
& \D \sum_{l=1}^{N}
\frac{\phi_{1l}\psi_{2l}}{\lambda -\lambda _l}
& \D \sum_{l=1}^{N}\frac{\phi_{1l}\psi_{3l}}
{\lambda -\lambda _l}
 \vspace{2mm}\\
\D \sum_{l=1}^{N}\frac{\phi_{2l}\psi_{1l}}
{\lambda-\lambda _l}
& \gamma _2+
\D  \sum_{l=1}^{N}\frac{\phi_{2l}\psi_{2l}}{\lambda-\lambda _l}
&\D  \sum_{l=1}^{N}\frac{\phi_{2l}\psi_{3l}}{\lambda -\lambda _l}
\vspace{2mm}
\\
\D \sum_{l=1}^{N}\frac{\phi_{3l}\psi_{1l}}{\lambda -\lambda _l}
& \D \sum_{l=1}^{N}\frac{\phi_{3l}\psi_{2l}}{\lambda-\lambda _l}
& \gamma _3+\D \sum_{l=1}^{N}\frac{\phi_{3l}\psi_{3l}}
{\lambda-\lambda _l}
\end{array}
\right).\label{eq:defofL(lambda)of3NWIEs} \ee 
Now if we set $\widetilde {U}(\la )=U(\tilde u,\la )$ and $ \widetilde {V}
(\la )= V(\tilde u,\la ),$ then  
it is easy to check the following Lax representations:
\be 
(L(\la ))_x=[\widetilde{U}(\la ),L(\la )],\ 
(L(\la ))_{t}=[
\widetilde{V}(\la ),L(\la )],
\label{eq:Laxrepresentationsofcfsof3NWIEs}
\ee 
for
the spatial constrained flow 
(\ref{eq:xcfof3NWIEs})
and the temporal constrained flow (\ref{eq:tcfof3NWIEs}),
respectively. 
These Lax representations 
will be applied to construct
the required integrals of motion for the Liouville 
integrability of two constrained flows in the next section. 

\section{$r$-matrix formulation
and Liouville integrability}
\setcounter{equation}{0}

Let us specify the
Poisson bracket: 
\be
\{f,g\}=
\sum_{i=1}^{3}\Bigl(\langle \frac{\partial f}{\partial \Psi_{i}},
\frac{\partial g}{\partial \Phi_{i}}\rangle -
\langle \frac{\partial f}{\partial \Phi_{i}},
\frac{\partial g}{\partial \Psi_{i}}\rangle \Bigr),\ 
f,g\in C^\infty (\R ^{6N}).
\label{eq:pbof3MWIEs}
\ee 

An $r$-matrix formulation can be directly shown for the Lax operator 
$L(\la )$ defined by 
(\ref{eq:defofL(lambda)of3NWIEs}).

\begin{theorem} The Lax operator $L(\lambda)$ given by (\ref{eq:defofL(lambda)of3NWIEs})
has the following $r$-matrix formulation
\be 
\{L(\lambda)\stackrel{\otimes}{,}
L(\mu)\}=\left[\frac{1}{\mu-\lambda}{\mathcal P},
L_{1}(\lambda)+L_{2}(\mu)\right] , \ 
 {\mathcal P}= \sum_{i,j=1}^3E_{ij}\otimes E_{ji},
\label{eq:rformulationofLaxoperatorof3NWIEs}\ee
where 
$L_1(\la )=L(\la )\otimes I_3,\ L_2(\la )=I_3\otimes L(\mu )$,
and 
\[
(E_{ij})_{kl}=\delta_{ik}\delta_{jl},\ 
\{L(\lambda)\stackrel{\otimes}{,}
L(\mu) \}_{ij,kl}=\{(L(\la ))_{ik},(L(\mu ))_{jl}\},\ 1\le i,j,k,l\le 3.
\ \]
\end{theorem}

First from the Lax representations in
(\ref{eq:Laxrepresentationsofcfsof3NWIEs}), we have 
\[ 
(\nu I_3-L(\la ))_x=[ \widetilde{U}(\la ),\nu I_3-L(\la )],\ 
(\nu I_3-L(\la ))_t=[ \widetilde{V}(\la ),\nu I_3-L(\la )],\ \]  
where $\nu $ is a constant,
and thus \cite{Tu-JPA1989}
\[ (\det (\nu I_3-L(\la )))_x=0,\ (\det (\nu I_3-L(\la )))_t=0.\]
Second from the $r$-matrix formulation (\ref{eq:rformulationofLaxoperatorof3NWIEs}), we have 
\cite{BabelonV-PLB1990} 
\[ \{tr L^k(\lambda ),tr L^l(\mu )\}=0,\  k,l\ge 1. \]
Expand the determinant of the matrix $\nu I_3-L(\la )$ as
\be \det(\nu I_3-L(\lambda))=\nu^3-{\mathcal F}_1(\lambda)\nu^2+{\mathcal F}_2(\lambda)\nu
-{\mathcal F}_3(\lambda),
\ee 
where by Newton's identities on 
elementary symmetric polynomials,
\[  \ba {rl} {\mathcal F}_1(\lambda)=&tr L(\lambda),\ {\mathcal F}_2(\lambda)
=\frac 1 2((tr L(\lambda))^2-
tr L^2(\lambda )),
\vspace{2mm}\\
{\mathcal F}_3(\lambda)=&\frac 1 6 (tr L(\lambda))^3+\frac 1 3tr L^3(\lambda)
-\frac 1 2(tr L(\lambda))tr L^2(\lambda).\ea \] 
Thus, it follows that 
\be  
({\mathcal F}_i(\la ))_x=0,\ ({\mathcal F}_i(\la ))_t=0,\ \textrm{and}\ 
\{{\mathcal F}_i(\la ),{\mathcal F}_j(\mu )\}=0,\ 1\le i,j\le 3. \ 
\label{eq:involutivepropertyofintegralsofmotion}\ee 
Further expand ${\mathcal F}_i(\la )$ as 
\be  {\mathcal F}_i(\la )=\sum_{l\ge 0}F_{il}\la ^{-l},\ 1\le i\le 3,\ee 
where obviously the $F_{i0}$'s are constants.
Then from the last equality in (\ref{eq:involutivepropertyofintegralsofmotion}), we have 
\be \{F_{il},F_{jk}\}=0,\ 1\le i,j\le 3,\ k,l\ge 0.\label{eq:ipofF_{il}F_{jk}}\ee 
The first two equalities in 
(\ref{eq:involutivepropertyofintegralsofmotion}) 
and the equality (\ref{eq:ipofF_{il}F_{jk}}) implies that 
two constrained flows (\ref{eq:xcfof3NWIEs})
and (\ref{eq:tcfof3NWIEs}) have the common involutive integrals of 
motion: $F_{il},\ 1\le i\le 3,\ l\ge 1.$
Now it is a direct computation to 
verify the following theorem on the Liouville integrability of the constrained flows
(\ref{eq:xcfof3NWIEs})
and (\ref{eq:tcfof3NWIEs}).

\begin{theorem} 
Two constrained flows (\ref{eq:xcfof3NWIEs})
and (\ref{eq:tcfof3NWIEs})
have the common involutive integrals of motion:
$F_{il},\ 1\le i\le 3,\ l\ge 1$,
of which the functions $F_{is},\ 1\le i\le 3,\ 1\le s\le N,$
are functionally independent
over a dense open subset of $\R ^{6N}$. Therefore, 
the constrained flows (\ref{eq:xcfof3NWIEs})
and (\ref{eq:tcfof3NWIEs}) are Liouville integrable.
\end{theorem}

\section{Involutive solutions}
\setcounter{equation}{0}

Since under the constraints on the potentials (\ref{eq:defoftildeu}),
the compatability condition of 
(\ref{eq:originalxpartofbinaryLaxsystems}) and 
(\ref{eq:originaltpartofbinaryLaxsystems})
is still the multi-wave interaction equations (\ref{eq:3MWIEs}),
solutions $(\Phi_i(x,t),\Psi_i(x,t))$ to the constrained flows (\ref{eq:xcfof3NWIEs}) and (\ref{eq:tcfof3NWIEs})
present solutions to the multi-wave interaction equations (\ref{eq:3MWIEs}):
\be 
u_{ij}(x,t)=\frac {\al _i-\al _j}{\gamma _i -\gamma  _j}
\langle\Phi_i(x,t) ,
\Psi_j(x,t)\rangle,
\ 1\le i\ne j\le 3.\label{eq:involutivesolutionsof3MWIEs}\ee 
This also shows the integrability by quadratures for the 
multi-wave interaction equations (\ref{eq:3MWIEs})
since $\Phi_i$ and $\Psi_i$ can be determined by quadratures.
On the other hand, it is easily found that 
under the Poisson bracket (\ref{eq:pbof3MWIEs}), two Hamiltonians $H^x$ and $H^t$ 
commute, i.e.,  
\be \{H^x,H^t\}= \sum_{i=1}^3 \Bigl(\langle \frac {\part H^x}{\part \Psi_i}
,\frac {\part H^t}{\part \Phi_i}\rangle - \langle
\frac {\part H^x}{\part \Phi_i},
\frac {\part H^t}{\part \Psi_i}
\rangle \Bigr) =0.\ee 
Hence, the above solutions given by (\ref{eq:involutivesolutionsof3MWIEs})
determine involutive solutions to the multi-wave interaction equations
(\ref{eq:3MWIEs}). If we denote two Hamiltonian flows of
(\ref{eq:xcfof3NWIEs}) and (\ref{eq:tcfof3NWIEs}) 
by $g_x^{H^x }$ and $g_{t}^{H^t}$ respectively,
then we have 
\[ \ba {rcl}    
u_{ij}(x,t)&=&
 \D \frac {\al _i-\al _j}{\gamma  _i -\gamma  _j}\langle g_x^{H^x}g_{t}^{H^t}\Phi_{i0},
g_x^{H^x}g_{t}^{H^t}\Psi_{j0}\rangle\vspace{1mm}  \\
&=&\D \frac {\al _i-\al _j}{\gamma  _i -\gamma  _j}\langle g_{t}^{H^t}g_x^{H^x}\Phi_{i0},
g_{t}^{H^t}g_x^{H^x}\Psi_{j0}\rangle,\ 1\le i\ne j\le 3,
\ea \]
where the initial values ${\Phi}_{i0}$ and ${\Psi}_{i0}$ 
of $\Phi_{i}$ and $\Psi_i$
can be taken as any arbitrary constant vectors of $\R ^N$.

Summing up, solutions of 
the constrained flows (\ref{eq:xcfof3NWIEs}) and (\ref{eq:tcfof3NWIEs}) 
lead to involutive solutions of the 
multi-wave interaction equations (\ref{eq:3MWIEs}). 
Moreover, such involutive solutions
show us 
(a) the richness of solutions of the multi-wave interaction equations (\ref{eq:3MWIEs}), and
(b) the integrability by quadratures for the multi-wave interaction equations (\ref{eq:3MWIEs}).

\section{Summary and conclusion}
\setcounter{equation}{0}
\label{sec:finalsectionofitaly00}

Binary Bargmann symmetry constraints in the continuous case decompose 
soliton equations (PDEs)
into constrained flows (ODEs).
The $r$-matrix formulation can be used to show 
the Liouville integrability of the constrained flows.
The resulting constraints on the potentials
give rise to involutive solutions to soliton equations
and thus show the integrability by quadratures for soliton equations.
The whole process of binary Bargmann symmetry constraints,
called binary nonlinearization, can be depicted as follows.
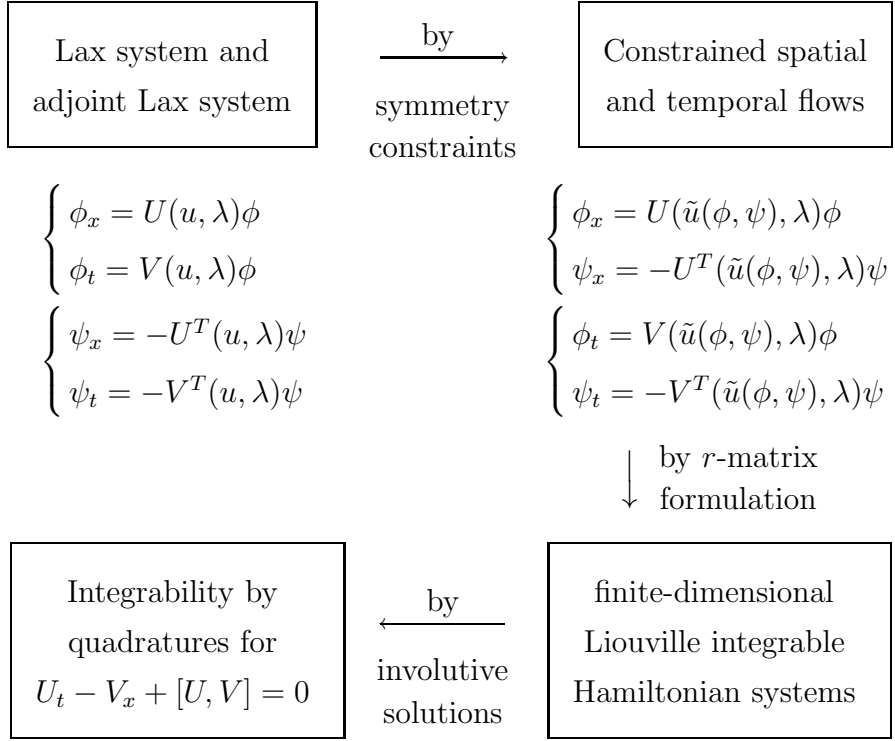
\[ \ba {ccc}
\fbox{$\ba {c}\textrm{Lax system and}\vspace{-1mm}\\ \textrm{adjoint Lax system}\ea $}\quad 
& 
\ba {c} \vspace{-2mm} \\
\stackrel {\stackrel {\textrm{by }}
{\bf 
{\frac{\quad }{\qquad \qquad } \hspace{-4.6mm}\longrightarrow }}
}{\ba {c} \textrm{symmetry} \vspace{-2mm}\\ \textrm{constraints}\ea }
\ea 
&
\quad \fbox{$\ba {c} 
\textrm{Constrained spatial}\vspace{-1mm}\\ \textrm{and temporal flows}
\ea  $}
\vspace{1mm}\\
\ba {l} 
\left \{\ba {l}  \phi_x=U(u,\la )\phi \vspace{0mm}\\
\phi_t=V(u,\la )\phi
  \ea \right. 
\vspace{1mm}\\
 \left \{\ba {l}  \psi_x=-U^T(u,\la )\psi \vspace{0mm}\\
\psi_t=-V^T(u,\la )\psi  \ea \right. 
\ea 
&  &
\ba {l}
\left\{ \ba {l} \phi_x=U(\tilde u (\phi,\psi),\la )\phi \vspace{0mm}\\
\psi_x=-U^T(\tilde u(\phi,\psi),\la )\psi  \ea \right. 
\vspace{1mm}\\
\left\{ \ba {l} \phi_t=V(\tilde u (\phi,\psi ),\la )\phi \vspace{0mm}\\
\psi_t=-V^T(\tilde u(\phi,\psi),\la )\psi  \ea \right. 
\ea 
\vspace{1mm}\\
& & \ba {c} {\bf \mid }\vspace{-4mm} \\ {\bf \downarrow }\ea 
\ \ba {c} \textrm{by $r$-matrix}\vspace{-2mm} \\ \textrm{formulation} \ea
\vspace{2mm}
\\
\fbox{$
\ba {c} \textrm{Integrability by} \vspace{-1mm}\\ \textrm{
quadratures for }\vspace{-1mm}\\
\textrm{$U_t-V_x+[U,V]=0$}
\ea $ }
&
\ba {c} \vspace{-2mm} \\
\stackrel
{\stackrel{\textrm{by}}{
{\bf \longleftarrow \hspace{-5mm}
\frac{\quad }{\qquad \qquad } }
}}
{\ba {c} \textrm{involutive} \vspace{-2mm}\\ \textrm{solutions}\ea }
\ea 
& 
\fbox{$
\ba {c}
 \textrm{finite-dimensional} \vspace{-1mm}\\
 \textrm{Liouville integrable} \vspace{-1mm}\\ \textrm{Hamiltonian systems}
\ea $ }
\ea 
\]

The multi-wave interaction equations 
(\ref{eq:3MWIEs}) have been taken as an illustrative example.
Binary Bargmann symmetry constraints (\ref{eq:bscof3MWIEs}), 
containing a set of arbitrary distinct constants $\gamma_1,\gamma_2,\gamma_3$,
were proposed for the multi-wave interaction equations (\ref{eq:3MWIEs}).
Two finite-dimensional Liouville integrable Hamiltonian systems 
(\ref{eq:Hsofxpartofcfs}) and (\ref{eq:Hsoftpartofcfs}})
resulted from the constrained flows determine involutive solutions to  
the multi-wave interaction equations (\ref{eq:3MWIEs}).
Our result with a special case 
$\Gamma =V_0$, i.e., $\textrm{diag}(\gamma_1,\gamma_2,\gamma_3)=\textrm{diag}(\beta_1,\beta_2,\beta_3)$, 
gives rise to all the results established in \cite{WuG-JMP1999}.

Of special interest in the study of binary symmetry constraints are to create new classical integrable Hamiltonian systems 
which supplement the known classes of classical integrable systems \cite{Perelomov-book1990}
and to expose the integrability by 
quadratures for soliton equations by using 
constrained flows \cite{MaF-book1996}.
We point out that high-order symmetry constraints with involved symmetries having 
degenerate Hamiltonians 
need particular consideration \cite{LiM-CSF2000}. 
A profound mathematical theory on binary symmetry constraints,
especially on Hamiltonian structures of constrained flows,
will be discussed elsewhere.

\vspace{3mm} 

\noindent {\bf Acknowledgments}

\noindent 
This work was supported by a grant from the Research Grants Council of 
Hong Kong 
Special Administrative Region, China (Project no. 9040466), and
a grant from the City University of Hong Kong (Project no. 7001041).
The author is also grateful to Prof. Z. X. Zhou for stimulating discussions.


\small 
\begin{thebibliography}{00}

\bibitem{Cao-SC1990} C. W. Cao, 
 Nonlinearization of the Lax system for AKNS hierarchy,
 {\em Sci. China Ser. A\/} 
 {\bf 33} (1990) 528--536.
 
\bibitem{CaoG-book1990} C. W. Cao and G. X. Geng,
Classical integrable systems generated through nonlinearization of
eigenvalue problems, In: C. H. Gu, Y. S. Li and G. Z. Tu, eds.,
{\em Nonlinear Physics\/}, Proceedings,
 Shanghai, 1989, (Springer-Verlag, Berlin, 1990) 
68--78. 

\bibitem{ZengL-AMS1990}
 Y. B. Zeng and Y. S. Li, 
 Three kinds of constraints of potential for KdV hierarchy,
 {\em Acta Math. Sinica (N.S.)\/}
 {\bf 6} (1990)
 257--272. 

\bibitem{MaS-PLA1994} W. X. Ma and W. Strampp, 
 An explicit symmetry constraint for the Lax pairs and the adjoint Lax  pairs of AKNS systems,
{\em Phys. Lett. A\/} {\bf 185} (1994) 277--286.

\bibitem{Ma-PA1995} W. X. Ma, 
 Symmetry constraint of MKdV equations by binary nonlinearization, {\em Physica
 A\/} {\bf 219} 
 (1995) 467--481.

\bibitem{MaFO-PA1996} 
 W. X. Ma, B. Fuchssteiner and W. Oevel, 
 A $3\times 3$ matrix spectral problem for AKNS hierarchy and its binary nonlinearization,
{\em Physica A\/} {\bf 233} (1996) 
 331--354.

\bibitem{EilbeckEKT-JPA1994}
 J. C. Eilbeck, V. Z. Enolskii, V. B. Kuznetsov and A. V. Tsiganov,
  Linear $r$-matrix algebra for classical separable systems,
{\em J. Phys. A: Math. Gen.\/} {\bf 27} 
  (1994) 567--578.
\bibitem{MaZ-krustal2000}
 W. X. Ma and Y. B. Zeng, 
Binary constrained flows and separation of variables for soliton equations, 
to appear in {\em 
Proceedings of the Conference
on Integrable Systems\/}
in celebration of Martin D. Kruskal's 75th Birthday,
 Adelaide, Australia, 2000.

\bibitem{MaF-book1996} W. X. Ma and B. Fuchssteiner, 
 Binary nonlinearization of Lax pairs,
 In: E. Alfinito, M. Boiti, L. Martina and F. Pempinelli, eds.,
 {\em Nonlinear Physics: Theory and
 Experiment\/}, Proceedings, Lecce, 1995 
 (World Sci. Publishing, River Edge, NJ, 1996) 217--224.

\bibitem{MaG-book1999} W. X. Ma and X. G. Geng, 
   B\"acklund transformations of soliton systems
    from symmetry constraints, to appear in 
   {\em Proceedings of the 
   AARMS-CRM Workshop on B\"acklund $\&$ Darboux Transformations:
   The Geometry of Soliton Theory\/}, Halifax, Canada, 1999.

\bibitem{Ma-CAM1997} W. X. Ma,  
 Binary nonlinearization for the Dirac systems,
{\em Chinese Ann. Math. Ser. B\/} {\bf 18} (1997) 
 79--88.

\bibitem{ZakhrovM-SPJEPTL1973} V. E. Zakharov and S. V. Manakov, 
 Resonant interaction of wave 
 packets in nonlinear media, {\em Sov. Phys. JETP Lett.\/} {\bf 18} (1973) 243--245.

\bibitem{Kaup-SAM1976} D. J. Kaup, 
 The three-wave interaction - a nondispersive phenomenon, {\em Stud. Appl. Math.\/} {\bf 55}   (1976) 9--44.

\bibitem{Tu-JPA1989}
G. Z. Tu, 
On Liouville integrability of zero-curvature equations and the Yang hierarchy, {\em J. Phys. A: Math. Gen.\/} {\bf 22} (1989) 
2375--2392. 

\bibitem{BabelonV-PLB1990} O. Babelon and C. M. Viallet, 
 Hamiltonian structures and Lax equations,
{\em Phys. Lett. B\/} {\bf 237} 
 (1990) 411--416. 

\bibitem{WuG-JMP1999} Y. T. Wu and G. X. Geng, 
  A finite-dimensional integrable system associated with the three-wave
  interaction equations, {\em J. Math. Phys.\/} {\bf 40} 
  (1999) 3409--3430.

\bibitem{Perelomov-book1990}
 A. M. Perelomov, {\em Integrable systems of classical mechanics and Lie algebras\/}, Vol. I.
(Birkh\"auser Verlag, Basel, 1990).

\bibitem{LiM-CSF2000}
Y. S. Li and W. X. Ma, Binary nonlinearization of AKNS spectral problem under higher-order symmetry constraints,
{\em Chaos, Solitons $\&$ Fractals\/} {\bf 11}
 (2000) 
697--710.

\end{thebibliography}
\end{document}